# Dose-efficient Automatic Differentiation for Ptychographic Reconstruction

LONGLONG WU[1,*], SHINJAE YOO[2], YONG S. CHU[3], XIAOJING HUANG[3,5], AND IAN K. ROBINSON[1,4,6]

[1]*Condensed Matter Physics and Materials Science Department, Brookhaven National Laboratory, Upton, NY 11973, USA*
[2]*Computational Science Initiative, Brookhaven National Laboratory, Upton, NY 11973, USA*
[3]*National Synchrotron Light Source II, Brookhaven National Laboratory, Upton, New York 11973, USA*
[4]*London Centre for Nanotechnology, University College London, London, WC1E 6BT, United Kingdom.*
[5]*xjhuang@bnl.gov*
[6]*irobinson@bnl.gov*
[*]*lwu@bnl.gov*

**Abstract:** Ptychography, as a powerful lensless imaging method, has become a popular member of the coherent diffractive imaging family over decades of development. The ability to utilize low-dose X-rays and/or fast scans offers a big advantage in a ptychographic measurement (for example, when measuring radiation-sensitive samples), but results in low-photon statistics, making the subsequent phase retrieval challenging. Here, we demonstrate a dose-efficient automatic differentiation framework for ptychographic reconstruction (DAP) at low-photon statistics and low overlap ratio. As no reciprocal space constraint is required in this DAP framework, the framework, based on various forward models, shows superior performance under these conditions. It effectively suppresses potential artifacts in the reconstructed images, especially for the inherent periodic artifact in a raster scan. We validate the effectiveness and robustness of this method using both simulated and measured datasets.

## 1. Introduction

As a powerful coherent diffraction imaging method[1-4], X-ray ptychography is now widely applied to image both material and biological structures by leveraging advanced synchrotron X-ray sources[5-7], X-ray free electron lasers[8-10], and high harmonic generation sources[11-13]. As a computational method for microscopic imaging, it enables the reconstruction of complete spatial information of the complex incident X-ray wavefront and high-resolution sample information from measured intensity-only coherent diffraction patterns. The intensity is a quantity with nonnegative real number values, as a detector collects a finite number of photons.

Because X-rays interact with matter in a variety of ways[14], the potential radiation-induced damage to the sample during a ptychographic measurement may limit its resolution and application. Meanwhile, the requirement of multiple coherent diffraction patterns with overlapping illumination in a ptychographic measurement is also time-consuming. Consequently, the utilization of low-dose X-rays and/or fast scans has become prevalent in a ptychographic measurement, albeit at the cost of lower photon statistics. Therefore, reliable performance of ptychographic reconstruction under low-photon statistics plays a critical role in the studies of radiation-sensitive materials, particularly biological structures where the ptychographic measurements need to be conducted using the lowest possible X-ray doses but to achieve a given resolution[15, 16]. It can also reduce acquisition time to speed up the ptychographic measurement while meeting the experimental requirements, such as time-resolved ptychographic study and ptychographic tomography measurement. Additionally, it can also facilitate the study of materials scattering weakly or, when attempting to reconstruct high-resolution images but only a few X-ray photons can be collected. However, ptychographic reconstruction at the low-photon statistics is a notoriously challenging task. Especially in the presence of shot noise, which varies in each experiment, necessitating different noise models for the recovery of the high-resolution complex-valued signals from intensity-only measurements further exacerbates the difficulty of the reconstruction under this condition.

Conventional ptychographic phase retrieval algorithms typically reconstruct the sample and probe information by retrieving the phase of complex far-field wave field[17]. These algorithms employ projection-based iterative methods, where the amplitude of the calculated far-field wavefront needs to be substituted with the measured one (*i.e.*, the reciprocal space constraint) at each iteration. Among these methods, gradient descent-based iterative approaches, such as the extended Ptychographic Iterative Engine (ePIE)[18], require an explicit gradient descent strategy for each optimizable parameter. Consequently, any change of the experimental condition and/or scattering model necessitates a manual re-derivation of the analytical expression for each optimizable parameter to obtain the corresponding gradient decent strategy, which is undesirable and makes the algorithms difficult to adapt to complex scattering models. An alternative approach to the "manually derived" gradient descent strategy for ptychographic reconstruction is the automatic differentiation (AD) method, which allows the automatic numerical calculation of the gradients of a loss function with respect to its optimizable parameters. Recently, AD-based ptychographic reconstruction methods have been successfully applied to experimental data[19, 20]. Its flexibility allows it to be easily adapted to the scattering model and experimental setup. However, its advantages have not yet been demonstrated for the low-photon statistical and low overlapping scenario, especially when complex scattering model is involved for example mixed state ptychographic reconstruction[19, 21, 22]. Meanwhile, similar to the conventional methods, the performances of the AD-based algorithms rely heavily on the initial parameters, notably such as the initial condition of the object and probe information, the choice of the batch size, and the learning rate for each optimizable parameter. Besides, a serious periodic artifact can be introduced from the periodicity of a raster ptychographic scan[23-25]. This is a long-standing problem in raster-scan ptychography that prevents the reconstruction of high-resolution sample information. Using a very high overlap ratio can suppress the artifact but will significantly increase the measurement time[23, 26, 27], which is not suitable for X-ray dose sensitive materials. Additionally, multimodal measurements (for example, simultaneous Ptychography and X-ray fluorescence)[7, 28, 29], for which raster scanning is effective, are continually growing in importance. These multimodal measurements will benefit greatly from improved analysis methods to reduce periodic artifacts in a raster scan.

In this work, we demonstrate a dose-efficient automatic differentiation framework for ptychographic reconstruction (DAP) under the low-photon statistical and low overlap conditions. Since there is no reciprocal space constraint (*i.e.,* the replacement of the calculated X-ray intensity with the measured one during the reconstruction) inside the method, based on this straightforward method, physics-constrained relationships, for example, the maximum likelihood estimation and the continuum property of materials, can be easily applied in the model to ensure the convergence of the algorithm. The robustness and efficiency of the proposed method are evaluated using both simulated and experimental ptychographic datasets, where the mixed state ptychographic reconstructions were applied by considering different noise models. When evaluating, the effect of the overlap ratio and the photon statistics on the existence of abovementioned periodic artifact is also investigated. The DAP was found to not only efficiently reconstruct high-quality images but also suppress the periodical artifacts under these low overlap and low photon statistics conditions. Furthermore, with the introduced variable-sized mini-bath optimization and autocorrelation initialization, the convergence has been significantly improved, resulting in higher-quality reconstructed results. As an experimental proof of concept, we expect this DAP approach will be widely adopted as a powerful and easy-to-adapt solution for ptychographic microscopes, especially when complex coherent scattering models are involved.

## 2. Results

### A. Model Description

In a forward X-ray ptychography experiment, the resulting complex exit wave field $\psi_i(\mathbf{r})$ can be generally expressed as[3, 17]:

$$\psi_i(\mathbf{r}) = P(\mathbf{r} - \mathbf{r}_i) \cdot O(\mathbf{r}), \tag{1}$$

where complex-valued object $O(\mathbf{r})$ interacts with a complex-valued X-ray probe beam $P(\mathbf{r})$ at position $\mathbf{r}_i$ to produce a complex-valued product $\psi_i(\mathbf{r})$. This "exit wavefront" propagates to the far field detector plane, approximated by a squared Fourier transform magnitude, describing the probability that the scattered X-ray photons from the propagated wavefront at position $\mathbf{r}_i$ can occur at the reciprocal-space vector q, such that:

$$D_i(\mathbf{q}) = |FT[\psi_i(\mathbf{r})]|^2, \tag{2}$$

where $FT$ represents the Fourier transform. Based on the incident X-ray dose, the experimentally recorded intensity-only coherent X-ray pattern $I_i(\mathbf{q})$, which is a quantity with nonnegative real number values, is different from $D_i(\mathbf{q})$ because of the statistical nature of photon counting, especially when the scattering signal is weak. The ultimate goal of a ptychographic measurement is to numerically retrieve the complex-valued object $O(\mathbf{r})$ and probe $P(\mathbf{r})$ using all the measured coherent diffraction patterns $I_i(\mathbf{q})$, such that each $I_i(\mathbf{q})$ can match the corresponding $D_i(\mathbf{q})$.

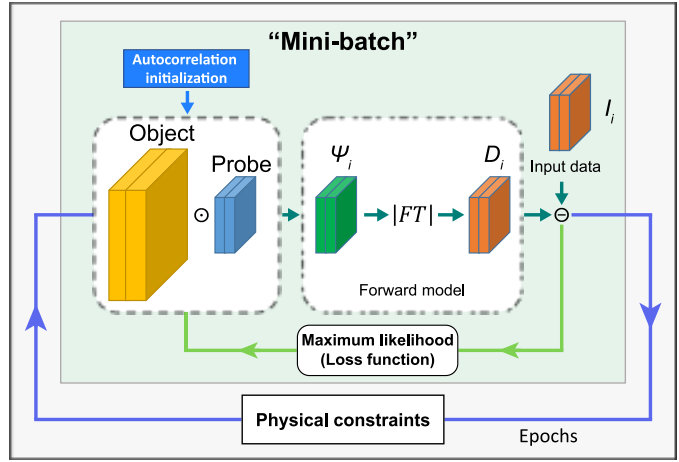

**Fig. 1.** Computational graph of the proposed dose-efficient automatic differentiation framework for ptychographic reconstruction. Input data (measured data - $I_i$) is in orange, and the backward propagation of the loss function is represented with the green line.

Figure 1 presents the computational graph of the proposed DAP approach for ptychographic reconstruction (see Supplement 1 for details). As with most gradient-based approaches, it is sensitive to the initial condition. Thus, since the inverse Fourier transform of far-field intensity is the autocorrelation of the exit wave, we propose an autocorrelation method to initialize the complex object $O(\mathbf{r})$ and X-ray probe $P(\mathbf{r})$ for the DAP algorithm. Briefly, the initial object is first estimated by $O(\mathbf{r}) = \beta_1 \cdot \left[ g(\mathbf{r}) \otimes \frac{\sum_{i=1}^{N} \varphi_i(\mathbf{r}+\mathbf{r}_i)}{\sum_{i=1}^{N} \Phi(\mathbf{r}+\mathbf{r}_i)} \cdot e^{\beta_2 \cdot g(\mathbf{r}) \otimes \frac{\sum_{i=1}^{N} \varphi_i(\mathbf{r}+\mathbf{r}_i)}{\sum_{i=1}^{N} \Phi(\mathbf{r}+\mathbf{r}_i)}} + \xi_1 \right]$, where $\varphi_i(\mathbf{r}) = |FT^{-1}(I_i)|$ and $\Phi(\mathbf{r}) = \left| FT^{-1}\left(\frac{1}{N}\sum_{i=1}^{N} I_i\right) \right|$. $\beta_1$ and $\beta_2$ are scale factor to normalize the initial object. $g(\mathbf{r})$ is a Gaussian smoothing kernel. The X-ray probe is initialized by $P(\mathbf{r}) = \frac{\zeta}{N}\left[\sum_{i=1}^{N}|FT^{-1}(\sqrt{I_i})| + \xi_2\right]$, where $\zeta$ is a scale factor to minimize the difference between the measured diffraction intensity and the calculated diffraction intensity calculated from the initialized object and probe. Here, $\xi_1$ and $\xi_2$ are additional Gaussian noise to avoid using the same initialization each time (see Supplement 1 and Fig. S1 and S2 for details). After initialization, based on the selected forward scattering model, the calculated X-ray scattering intensity $D_i(\mathbf{q})$ will be compared with its corresponding experimental recorded $I_i(\mathbf{q})$ through a loss function to optimize $O(\mathbf{r})$ and $P(\mathbf{r})$.

As a nonlinear optimization problem, we adopt a "mini-batch" gradient descent strategy to find the minimum of a loss function, where a subset of samples from the input dataset (*i.e.*, less than the full dataset) is used at each iteration until all the measured coherent diffraction patterns have been used. During each iteration, the target variables will be updated with each input subset. However, it should also be noted that, unlike the conventional mini-batch gradient descent strategy, where a fixed mini-batch size is applied during the optimization, the mini-batch size in our proposed DAP generally increases as the epochs increase. This approach was found to significantly improve the convergence of the algorithm (see Supplement Fig. S3 for more details).

As the mini-batch size varies in our proposed DAP during the optimization, in a particular case where its mini-batch size is set to 1, the DAP approach is similar to the traditional gradient descent based iterative methods such as ePIE. However, different from these algorithms, where a sequential update for each optimizable parameter needs to be made after calculating the gradients at each illuminated position, the updates in DAP are made for each position within a mini-batch, allowing for parallel calculation. Furthermore, when using the total number of the input datasets as the mini-batch size, the DAP has the most stable gradient for each optimizable parameter. Thus, variation of the mini-batch size during optimization will affect the uncertainty in the gradient for each parameter. The noise gradient is helpful for jumping out of local minima while a stable gradient will benefit the convergence of the reconstruction. In each epoch, once all the measured coherent diffraction patterns have been used, the X-ray probe $P$ will be recentered to remove any global translation ambiguity, and its mean phase will be set to zero. In the meantime, the complex object $O$ will be renormalized to remove any uncertainty of the scaling effect between the object $O$ and X-ray probe $P$. Additionally, the phase range of object $O$ will also be constrained if a range is set.

The correct selection of the learning rates for each optimizable parameter is important for a successful reconstruction. High values of learning rates can make the optimization scheme unstable and the divergence of the optimizable parameters, while low values of learning rates can impede the convergence of the optimizable parameters, for example, slow convergence. For the proposed DAP, the initial learning rate for the object is generally around 0.15 as the initialized object $O$ is normalized. However, for the X-ray probe, its initial learning rate is adjusted based on the mean value of its amplitude. During the optimization, the learning rate for each parameter will be dynamically reduced by its corresponding scheduler using the loss metrics quantity when no improvement is seen for a 'patience' number of epochs. Here, we use Adaptive Moment Estimation (Adam) as the optimizer to update the underlying variables, such as $O$ and $P$, which is a modification of the RMSProp optimizer, using moving averages on both the gradient and the second moment of the gradient[20]. To make the optimization process more flexible, each optimizable parameter will be optimized with its independent optimizer so that a different schedule can be set up for each parameter.

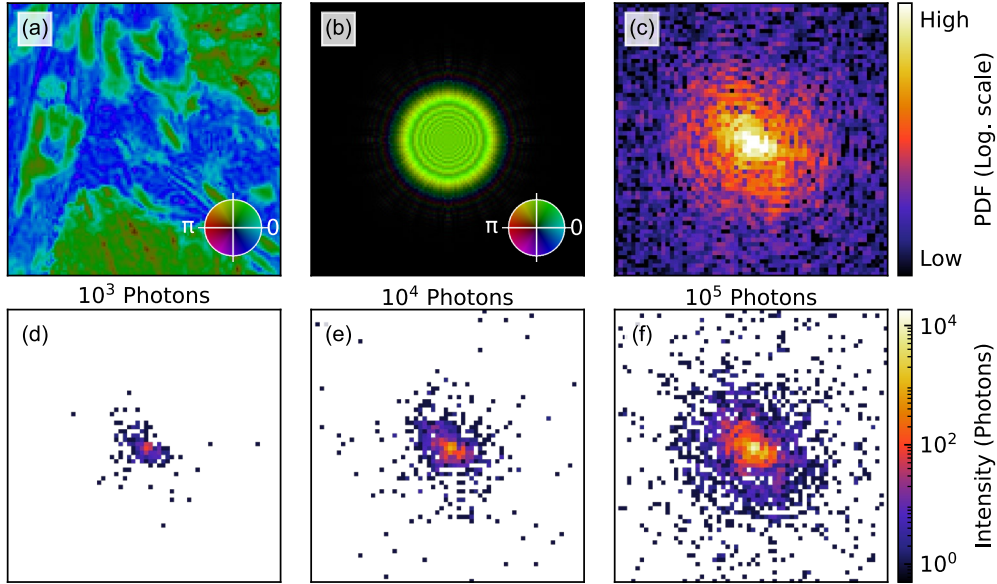

**Fig. 2.** Effect of photon statistics on the experimental coherent diffraction pattern. (a) Test complex sample in HSV format. (b) X-ray probe used in the optical simulation. (c) Calculated theoretical coherent X-ray diffraction pattern using (a) and (b). Simulated X-ray diffraction patterns with the amount of scattered photons of (d) $10^3$, (e) $10^4$ and (f) $10^5$.

The loss function is the central feature of an optimization process. The measured coherent diffraction intensity is generally a quantity with nonnegative real number value. Due to finite photon counting, the recorded coherent diffraction intensity will differ from its corresponding scattering probability, especially when the scattering signal is weak, regardless of the external noise. Figure 2 demonstrates this effect using a numerical simulation, where the implementation is based on an acceptance-rejection method (see Supplemental 1 for details)[30, 31]. The simulations were performed using the test object given in Fig. 2(a), based on a database photograph[32], and the X-ray probe given in Fig. 2(b). Here, the X-ray probe is obtained by propagating a circular shape to a certain distance using the angular spectrum method. Figure 2(c) presents the corresponding theoretical scattering probability (*i.e.,* probability density function, PDF), which is proportional to the modulus square of the Fourier transform of the test object weighted by the X-ray probe. As presented in Fig. 2(d)-(f), when the number of scattered photons is increased from $10^3$ to $10^6$, the effect of photon statistic is reduced but is still clearly visible. The difference between simulated coherent diffraction and the theoretical one becomes more apparent as the X-ray photon statistic is low. In consequence, for a ptychographic reconstruction, a more accurate scattering model at low-photon statistics can reduce the number of unknowns that need to be estimated and, hence, should produce higher reconstruction quality.

In X-ray ptychographic imaging, the recorded intensity of coherent diffraction patterns is typically related to the number of photons that strike a detector pixel within a fixed exposure time. The collected X-ray photons at the detector are random in nature. The standard picture of this photon counting statistic shows that measured pixel counter recordings (or intensity), $I_i(\mathbf{q})$, follow the Poisson probability distribution function (see Supplemental 1 for details). Thus, considering the negative log-likelihood minimization for a Poisson distribution, the corresponding loss function for one coherent diffraction pattern in a ptychographic reconstruction can be expressed as:

$$\ell_{\mathcal{P},i}(\mathbf{q}) = \frac{1}{J}\sum_{J} D_i(\mathbf{q}) - I_i(\mathbf{q}) + I_i(\mathbf{q}) \log\left[\frac{I_i(\mathbf{q})}{D_i(\mathbf{q})}\right], \quad (3)$$

where $J$ is the number of pixels of the diffraction pattern. Here, an extra constant was introduced to the above function to make the Poisson log-likelihood estimation non-negative (see Supplement 1 for details). Further, when dealing with counting statistics, if the measured data are corrupted by an additive thermal noise to the square root (or amplitude) of the expected intensity[33, 34], where the noise can be approximated by an asymptotic form Gaussian counting model, the loss function, in this case, can be given as:

$$\ell_{\mathcal{G},i}(\mathbf{q}) = \frac{1}{J}\sum_{J}\left[D_i^{1/2}(\mathbf{q}) - I_i^{1/2}(\mathbf{q})\right]^2. \quad (4)$$

Additionally, if the Gaussian noise is additive to the expected intensity directly with its corresponding variance approximated by the measured intensity, the loss function in this situation can be expressed as[34, 35]:

$$\ell_{\mathcal{R},i}(\mathbf{q}) = \frac{1}{2J}\sum_{I_i(\mathbf{q})\neq 0}\left[\frac{D_i(\mathbf{q}) - I_i(\mathbf{q})}{I_i^{1/2}(\mathbf{q})}\right]^2 + \frac{1}{J}\sum_{I_i(\mathbf{q})=0} D_i(\mathbf{q}), \quad (5)$$

where the last part in the above equation is introduced to constrain the zero intensity for reducing noisy solutions.

In a forward ptychographic measurement, the finite X-ray photon statistics and inevitable noise in the measured coherent X-ray diffraction patterns may lead to artifacts in the reconstructed object. Especially under low-photon statistics, it will become more significant. However, it can usually be assumed that the projected refractive index of the sample is continuous in a ptychographic measurement, and the corresponding measured amplitude and phase information of the sample has a similar distribution. Thus, to minimize the potential artifacts, one popular idea is to introduce an additional regularization total variation (TV)[36] term to the negative log-likelihood function by penalizing variations in the complex object $O$. However, simply minimizing the TV of an image could lead to blurring[34, 37-40] and the best quality of the reconstructed image may not be achieved. Therefore, inspired by several image-denoising works, an adaptive $L_p$-norm based TV (ATV, *i.e.,* $L_p$ norm to the power of $q$) denoising is applied in our proposed DAP, which is written as:

$$ATV(O) = \frac{1}{HV}\sum_{h=1}^{H}\sum_{v=1}^{V}(|\nabla_x O|^p + |\nabla_y O|^p + \epsilon)^{\frac{q}{p}}, \quad (6)$$

where $\nabla_x$ and $\nabla_y$ denote the finite difference operations along the horizontal and vertical direction of the complex object $O$, respectively. H and V are the corresponding quantity of pixels along the horizontal and vertical direction of the complex object $O$, respectively. $\epsilon$ is an exceedingly small constant set to prevent singular gradient error. Unlike the classical TV model, the proposed ATV doesn't blindly pursue smoothness. It is adaptive and can be adjusted for every pixel of the reconstructed image, which can preserve the subtle features of the reconstructed object $O$ better. Especially, when $p = 1$ and $q = 1$, it becomes the classical $L_1$ TV, where it may treat noise as edges and generate false edges, giving a "ladder" effect. When $p = 2$ and $q = 1$, it becomes $L_2$ TV denoising model, which can prevent the ladder effect without generating false edges. Besides, for the ATV, the other combination of the $p$-and $q$-value also impacts different images differently (see Fig. S4 and S5 for more details).

Finally, the loss function used in the proposed DAP for a single state ptychographic reconstruction is a combination of maximum likelihood estimation and ATV, averaged over all the probe positions in *l*th mini-batch, which is given as:

$$\mathcal{L}_{\alpha,l} = \frac{1}{L} \sum_{i \in \Omega} \ell_{\alpha,i}(D_i, I_i) + \gamma \cdot ATV(O), \qquad (7)$$

where $\Omega$ contains the indices of coherent diffraction patterns in a mini-batch with the size of $L$, which generally increases as the epoch increases. Here, the subscript $\alpha$ stands for $\mathcal{P}$, $\mathcal{G}$, and $\mathcal{R}$ (*i.e.*, different statistical models under consideration). $\gamma$ is a coefficient that is dynamically changed to keep the ratio between maximum likelihood estimation and ATV fixed. Different from using a constant, we found this dynamical adjustment of the coefficient $\gamma$ can significantly enhance the convergence of the DAP algorithm and allow one to have large learning rates during the reconstruction. Since $\gamma$ is dynamically adjusted to keep the ratio fixed, at the beginning of the reconstruction, the ATV will have a strong effect on the object. However, as the loss decreases, the effect of ATV will decrease. This optimization seeks a solution that fits the maximum likelihood model but also has a limited TV for the reconstructed object.

## B. Performance on Simulated Data

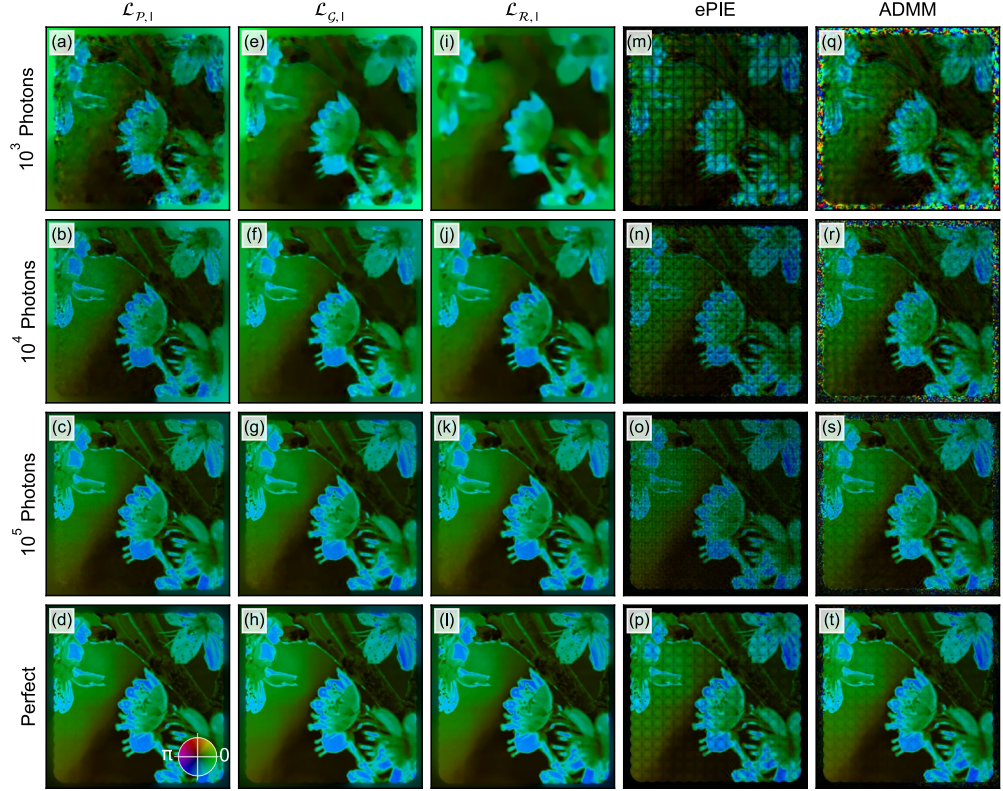

**Fig. 3.** Performance of the DAP on simulated ptychographic datasets with different photon statistics. (a)-(d) obtained results using the $\mathcal{L}_{\mathcal{P},l}$ loss function. (e-h) Using the $\mathcal{L}_{\mathcal{G},l}$ loss function. (i)-(l) Using the $\mathcal{L}_{\mathcal{R},l}$ loss function. (m)-(p) Corresponding results using the conventional ePIE algorithm. (q)-(t) Corresponding results using the ADMM algorithm. Here, the scanning overlap ratio is 50% for all images.

In an attempt to highlight the performance of our proposed DAP approach and demonstrate the intrinsic merit of different loss functions on the final reconstructed results, we first performed a numerical study based on Eq. 7. Based on the introduced acceptance-rejection method (see Supplement 1 for details), the ptychographic datasets with different photon counts were simulated using the X-ray probe beam from Fig. 2b with an array size of 64 × 64 pixels. The raster grid was obtained by translating a test photographic object with a total phase range of π radians. The step sizes in the horizontal and vertical direction are both 18 pixels. The corresponding overlap ratio is 50%, defined by the ratio of the scanning step size to the diameter of the probe (see Supplement Fig. S6 for details). A total of 225 diffraction patterns with an array size of 15×15 was generated for each ptychographic dataset. As the X-ray probe scans on the sample, each diffraction pattern will have a different amount of scattered photons, which was determined by the corresponding optical properties of the illuminated region of the sample. Since there is no noise added, the simulated X-ray intensity in each coherent diffraction pattern obeys the intrinsic Poisson statistics. As shown in Fig. 3, the DAP algorithm was applied to four different simulated ptychographic datasets from the same sample with different amounts of scattered photons. Here, the labels of $10^3$, $10^4$, and $10^5$ photons in Fig. 3 indicate the number of scattered photons of a coherent diffraction pattern, which is the maximum among all the coherent diffraction patterns in one ptychographic dataset (see Supplemental Fig. S7). They are also equal to the amount of the incident photons when assuming scattering efficiency of the material equals to 1. For minimization, some loss functions may undergo strong degradation when the initialization of the algorithm, for example the initial object, is far from the final solution. Thus, during the reconstructions, to reduce the impact of the optimization method on the reconstructed results, we use the $L_{G,l}$ loss function in the first 100 epochs to get a quick estimation and then switch to the target loss function for further minimization with another 100 epochs (see Supplement Fig. S8 for details). An estimated scale factor is also applied to the following loss function to reduce the effect of sudden gradient difference when switching (see Appendix A for the estimations).

As shown in Fig. 3(a-l), the DAP approach can achieve a decent reconstruction result with ~$10^3$ photons. However, in sharp contrast, the conventional ePIE algorithm suffers from the periodic artifact in both the reconstructed object and X-ray probe, making it difficult to identify the subtle features inside the object as shown in Fig. 3(m-p). Additionally, by utilizing the Alternating Direction Method of Multiplier (ADMM) algorithm[41], similar behaviors are also observed, shown in Fig. 3(q-m) [also see Supplemental Fig. S9-14 for more comparisons with more different algorithms using different overlap ratio, *i.e.,* 50%, 40%, 30%, 25%]. By further comparing the images in Fig. 3(a-l) obtained with different models, we find the feature in Fig. 3(a) is the sharpest, and Fig. 3(i) shows a relatively blurred image. When the amount of scattered photons is increased to ~$10^4$, the images obtained by the DAP approach still show the best-reconstructed results, compared with the corresponding result obtained with ePIE. The features in Fig. 3(b) are still better reconstructed. As it is further increased to ~$10^5$, the difference between these images obtained with DAP using different loss functions becomes insignificant. However, the ePIE algorithm still shows its weakness in the periodic artifact. Even when the ideal diffraction pattern is used, there is no trend seen this effect can be mitigated. To have a quantitative comparison, we further use the complex Pearson Correlation Coefficient (see Appendix A for its definition) to evaluate the quality of the reconstructed object from different reconstruction methods used in this paper. As presented in Fig. S15, by comparing the reconstructed object with its ground truth, it further confirms that DAP shows best performance. This is due to that most of conventional iterative methods compared in this paper generally don't have an explicit smoothness constraint like ATV for the object. This makes it difficult for the method to get rid of the periodic artifact arising from the symmetric scan trajectory. However, with ATV regularization and its dynamical adjustment applied in the DAP, the periodic artifact can be significantly suppressed. As given by Eq. 6 and 7, the ATV is a measure of the complexity of the object with respect to its spatial variation using both real and imaginary parts of the gradient of the object. During the optimization, any sudden change of the gradient in the object will increase the value of ATV. By changing the coefficient $\gamma$, we can control the penalty term in Eq. 7. The higher the $\gamma$ coefficient, the more it reduces the variation of the object's gradient, resulting in a smoothed object. As the maximum likelihood estimation is generally large at the beginning of an optimization, utilizing this dynamical adjustment strategy, the ATV will have a strong effect on the object. As the algorithm is converging, the effect from ATV is then reduced by tuning the of $\gamma$. With this strategy, the ultimate convergence of the algorithm is preserved, while keeping the periodic artifact mitigated (see Fig. S16 for details). With the above comparison, the proposed DAP shows much better performance than the conventional algorithms. Even when the X-ray photon statistic is low, the DAP not only can give a well-distinguishable object image but also can suppress the factorization artifact that degrades the ptychography. Additionally, it should be noted that for different photon statistics, each noise model behaves differently.

## C. Application to Experimental Data

Because the experimental illumination condition is more complicated by the presence of multiple optical modes[18, 42], we further applied the DAP approach to experimental ptychographic datasets to demonstrate its capability. Assuming that the physical object and X-ray probe can be effectively modeled by $M$ independent object states and $N$ independent probe states (see Supplement 1 for more details), the corresponding loss function for this mixed state ptychographic reconstruction can be written as:

$$\mathcal{L}_{\alpha,l} = \frac{1}{LMN} \sum_{i \in \Omega} \sum_{m=1}^{M} \sum_{n=1}^{N} \ell_{\alpha,i}[D_i^{(m,n)}, I_i^{(m,n)}] + \frac{\gamma}{M} \sum_{m=1}^{M} ATV(O_m), \qquad (8)$$

where $D_i^{(m,n)}$ is the calculated coherent diffraction pattern from the far field wavefront of the $m$-th state of the object and $n$-th state of the X-ray probe at the scanning position $\mathbf{r}_i$ and $I_i^{(m,n)}$ is the corresponding experimental pattern with $I_i^{(m,n)} = I_i \frac{D_i^{(m,n)}}{\sum_{m,n} D_i^{(m,n)}}$. Particularly, when $M = 1$ and $N = 1$, Eq. 8 can be simplified to Eq. 7. The ptychographic experiments were performed at the Hard X-ray Nanoprobe (HXN) beamline, National Synchrotron Light Source II (NSLS-II) using focused X-ray beams from the Multilayer Laue Lenses (MLLs) and Fresnel Zone Plate (FZP), respectively (see Appendix A for more details). The incident X-ray beam filtered by a double crystal Si (111) monochromator was pre-focused at the secondary source aperture plane, which is about 15 m in front of the nano-focusing optics. A Siemens Star test pattern made of Au was used to acquire ptychographic datasets with different exposure times. Each scan was taken in a fly-scan mode. For the MLLs, the energy of the incident X-ray beam was 15 keV, and it was 10 keV for the FZP.

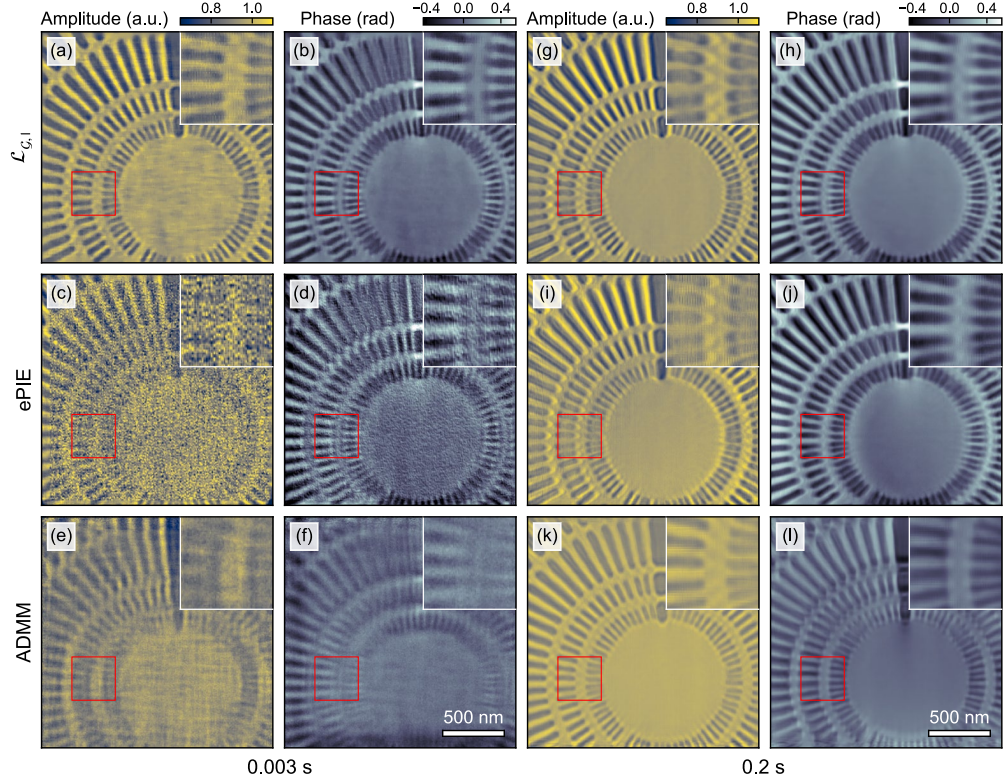

**Fig. 4.** Experimental ptychographic reconstruction of a Siemens star object with MLLs using different exposure times. (a) Reconstructed amplitude using the proposed DAP with 0.003 second exposure. (b) Corresponding reconstructed phase. (c) Reconstructed amplitude using ePIE. (d) Corresponding reconstructed phase. (e) Reconstructed amplitude using ADMM. (f) Corresponding reconstructed phase. (g) Reconstructed amplitude using our proposed DAP with 0.2 second exposure. (h) Corresponding reconstructed phase. (i) Reconstructed amplitude using ePIE. (j) Corresponding reconstructed phase. (k) Reconstructed amplitude using ADMM. (l) Corresponding reconstructed phase. Here, the insets show enlarged views of the red-boxed region.

Figure 4 shows the comparison of the reconstructed results using our proposed DAP with $\mathcal{L}_{\mathcal{G},l}$ loss function and the conventional ePIE and ADMM algorithms from the acquired experimental datasets (also see Fig. S17 for its performance on simulated data). Both datasets are obtained with a defocus X-ray beam from the MLLs. When performing the reconstructions, one object state and two probe states were applied (see Fig. S18 for obtained probes). As presented in Fig. 4(a-f), the used experimental ptychographic dataset was measured with 0.003 s exposure time for each pattern, where the average amount of scattered photon for each coherent pattern is only ~5778 (*i.e.,* ~0.12 photons per pixel). The proposed DAP yields high-quality reconstruction where the small features can still be well distinguished at this low-photon statistics condition. However, in contrast, the corresponding reconstructed images from ePIE and ADMM algorithms show visible noisy features, especially in the obtained amplitude information of the Au Siemens star. Further, Fig. 4(g-l) presented the reconstructed results, where the dataset was obtained with 0.2 s exposure time. The corresponding number of scattered photons per pattern is ~385175 (*i.e.,* ~7.96 photons per pixel). Still, the reconstructions from the DAP algorithm present a much better resolution (see Fig. S19 for the calculated phase retrieval transfer function). Thus, the proposed DAP can achieve a decent resolution under low-photon statistics, which can greatly facilitate low dose and/or fast scan ptychographic measurement.

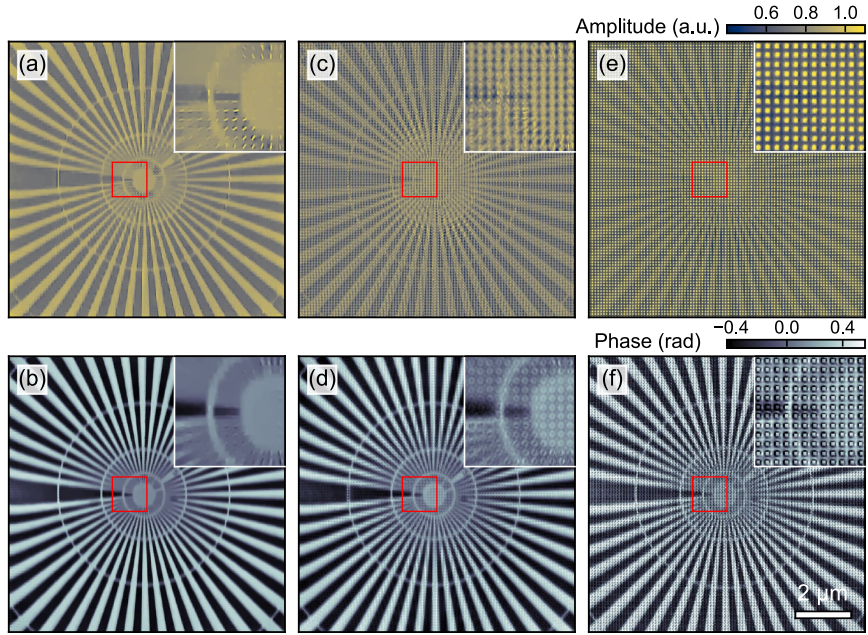

**Fig. 5.** Experimental ptychographic reconstruction of a Siemens star object with FZP. (a) Reconstructed amplitude using the proposed DAP. (b) Corresponding reconstructed phase. (c) Reconstructed amplitude using ePIE algorithm. (d) Corresponding reconstructed phase. (e) Reconstructed amplitude using ADMM algorithm. (f) Corresponding reconstructed phase. Here, the insets show enlarged views of the red-boxed region.

Furthermore, as various noises in experimental data are mixed with the diffracted signals on the detector, some of them can transform into artifacts in reconstructed images. Meanwhile, when the overlap ratio is low, the periodic artifacts arising from the factorization effect between the object and the probe, is a well-known problem in raster scan ptychography[24]. Ptychographic reconstruction under these realities becomes a challenging task. Since the traditional iterative reconstruction methods suffer from periodic artifact, several approaches have been proposed to remove or suppress the grid pathology in raster-scan ptychography[23, 43, 44]. However, these methods generally require prior knowledge of the experiment. For example, the size of an X-ray probe is required to estimate its corresponding support, which is not suitable for a highly structured X-ray beam. To further demonstrate the capability of the proposed DAP, Figure 5 shows the reconstructed results using the dataset measured from the FZP with a very low overlap ratio. When performing the reconstructions, one object state and four probe states were applied for DAP, ePIE, and ADMM (see Fig. S20 for the obtained probes). The results from DAP were obtained with $\mathcal{L}_{\mathcal{G},l}$ loss function. As shown in Fig. 5(a-b), the spokes can be well recognized. However, as shown in Fig. 5(c-f), the reconstructed images from the ePIE and ADMM algorithms were destroyed by the periodic artifact. Thus, with the proposed DAP, the periodical artifacts are seen to be significantly suppressed, which further endorses the advantage of the DAP algorithm. Additionally, we also applied these three algorithms to another ptychographic dataset with different scanning step sizes, obtained with one sub-micron gold crystal from a dewetted gold film. As shown in Fig. S21, the periodic artifacts are still presented in these reconstructed images from ePIE and ADMM. However, the artifact is avoided by DAP, which further endorses the advantage of our proposed method.

### 3. Discussion

As a straightforward optimization method, one significant distinction between our DAP algorithm and conventional methods is that no reciprocal space constraint (*i.e.,* using the measure X-ray intensity to substitute the calculated one during reconstruction) is applied during the optimization. The gradient is automatically numerically computed for the DAP method and no explicit knowledge of the gradient descent strategy for each optimizable parameter is needed. Thus, the DAP has its unique advantage to incorporate a more complex scattering model of a ptychographic experiment by a simple change of the loss function, especially when applied to some scattering models where the corresponding gradient descent strategy cannot be manually derived. Conversely, exact gradient descent knowledge is required for the conventional iteration methods. Also, as demonstrated, in low-photon counting ptychographic imaging experiments, the correct choice of the noise model plays a crucial role in the reconstruction of high-quality images. Currently, most of the ptychographic experiments are conducted using fly scans. We believe the proposed DAP will perform better by further considering the continuous movement of the X-ray beam on the sample and combining it with the other physical processes, for example, partial coherence. The extra constraint on the X-ray probe may also improve DAP's performance. However, these models will consume more computational resources where a balance between reconstruction accuracy and computational cost may need to be considered. As the DAP can significantly mitigate the periodic artifacts, it allows the ptychographic measurement under low overlap ratios using a simple raster grid scan which can help the related measurement a lot. Another important feature of our DAP method is its variable-size mini-batch, which interleaves the advantages of the ePIE and DM traditional methods. Additionally, within each mini-batch, the maximum likelihood estimation for each diffraction pattern is independent. Therefore, the proposed DAP can be easily adapted for parallel computation, reducing the reconstruction time. As the mini-batch size increases, the computational time will be significantly reduced. One may need to tune the parameter for the ATV and variable-size minibatch sequence to reconstruct better results when different data is applied. In the future, DAP's performance on probe position refinement, multi-slice ptychography, X-ray intensity fluctuation can be further explored.

## 4. Conclusion

In this paper, we have demonstrated a Dose-efficient Automatic differentiation framework for Ptychographic reconstruction (DAP) by considering various noise models. The DAP can converge faster with higher accuracy over current state-of-the-art algorithms. As there is no requirement for the analytical expression of the gradient descent strategy for each optimizable parameter, the DAP can greatly simplify the design of a reconstruction process and allow the incorporation of the different complex scattering models, for example, the mixed state ptychographic reconstruction with different noise models as we demonstrated in the paper. Meanwhile, by varying the mini-batch size during the reconstruction, the method can interleave the advantages of the conventional methods, such as ePIE and DM, and can achieve much better resolution under low-photon statistics. Especially, as demonstrated in the paper, the existence of the generalized total variation in the loss function and its dynamical adjustment can greatly enhance the convergence of the reconstruction and mitigate the long-standing periodic artifact for conventional methods when a raster scan grid with large step size is used. The inclusion of adaptive total-variation constraints will allow our proposed DAP to perform well in future applications with sparse or noisy data.

**APPENDIX A: METHODS**

## 1. Ptychography measurements

The Ptychography experiment was performed at the hard X-ray nanoprobe beamline (HXN) of National Synchrotron Light Source II, Brookhaven National Laboratory. The microscope sits about 15 m downstream from the secondary source aperture, and a Fresnel zone plate (Applied Nanotools Inc.) with 30-nm outmost zone width or Multilayer Laue Lenses (MLLs) was used to focus the beam to a nano spot. After the nanofocusing optics, there was an order-sorting-aperture blocking all undesired background signals. The Au sample was mounted inside the specially designed microscope with high stiffness and thermal stability. The incident X-ray beam energy for FZP is 9 KeV, and it is 15 KeV for MLLs. A pixel-array detector (Merlin, Quantum Detectors) was positioned 0.5 m downstream for FZP to record the transmitted far-field diffraction patterns. It was positioned 1 m for MLLs. For FZP, we performed a 2D raster grid scan with a range of $10 \times 10$ μm² um. The scanning step size is 100 nm in each direction. The corresponding diffract pattern size is $128 \times 128$ pixels and there are $101 \times 101$ frames inside the dataset. For MLLs, 2D raster grid scans with a range of $2 \times 2$ μm² were performed. The corresponding scanning step size is 10 nm in each direction. The size of the far-field diffraction pattern is $220 \times 220$ pixels and there are $201 \times 201$ frames inside each ptychographic dataset.

## 2. DAP implementation and ptychographic reconstruction

The DAP algorithm was implemented based on the PyTorch package (*i.e.,* version 2.1.2), where the gradient calculation is obtained using Wirtinger calculus for complex-valued array. When doing the reconstruction, the abovementioned autocorrelation approach will be first used to initialize the complex object and probe for the DAP algorithm. Then, to minimize the difference between the experimental diffraction pattern and the calculated diffraction pattern $D_{i,\text{init}}$, using the initialized object and probe, the scale factor $\zeta$ of the X-ray probe will be optimized with a least square fitting, *i.e.,* $\zeta = \underset{\zeta}{\text{argmin}}\left(\sum_{i=1}^{N}\sqrt{I_i} - \sqrt{D_{i,\text{init}}}\right)^2$. During the reconstruction, the ptychographic reconstructions were completed using the Adam optimizer. The learning rate is initialized to 0.15 for the object, and it is adjusted for the probe based on the mean of its amplitude. Both learning rates are dynamically reduced by the scheduler using the loss metrics quantity when no improvement is seen for a 'patience' number of epochs. The mini-batch size generally increases as the reconstruction epoch increases. For each epoch, based on the corresponding mini-batch size the experimental coherent diffraction will be divided randomly into different groups. When switching from one loss function to another, due to

$$\lim_{D_i(\mathbf{q}) \to I_i(\mathbf{q})} \frac{\left[D_i^{1/2}(\mathbf{q}) - I_i^{1/2}(\mathbf{q})\right]^2}{D_i(\mathbf{q}) - I_i(\mathbf{q}) + I_i(\mathbf{q}) \log\left[\frac{I_i(\mathbf{q})}{D_i(\mathbf{q})}\right]} = 1/2 \quad , \quad \lim_{D_i(\mathbf{q}) \to I_i(\mathbf{q})} \frac{D_i(\mathbf{q}) - I_i(\mathbf{q}) + I_i(\mathbf{q}) \log\left[\frac{I_i(\mathbf{q})}{D_i(\mathbf{q})}\right]}{\frac{1}{2}\left[\frac{D_i(\mathbf{q}) - I_i(\mathbf{q})}{I_i^{1/2}(\mathbf{q})}\right]^2} = 1 \quad \text{and}$$

$$\lim_{D_i(\mathbf{q}) \to I_i(\mathbf{q})} \frac{\frac{1}{2}\left[\frac{D_i(\mathbf{q}) - I_i(\mathbf{q})}{I_i^{1/2}(\mathbf{q})}\right]^2}{\left[D_i^{1/2}(\mathbf{q}) - I_i^{1/2}(\mathbf{q})\right]^2} = 2$$, a scale factor for the following loss function will be applied to avoid the sudden jump of the gradient, which was found that this can make the reconstruction more stable. When doing the reconstruction with ATV constraint, we set $p = 2$ and $q = 1$ for both simulated and experimental ptychographic datasets. The ptychographic reconstructions with conventional iterative phase-retrieval methods were completed using GPU-accelerated codes with Python[6, 7]. The complex Pearson Correlation Coefficient (cPCC) is used as a quantitative metric to evaluate the quality of the reconstructed complex object, defined as:

$$\text{cPCC}(x_1, x_2) = \frac{\sum_{s=1}^{S}(x_{1,s} - \overline{x_1})(x_{2,s}^* - \overline{x_2^*})}{\sqrt{\sum_{s=1}^{S}(x_{1,s} - \overline{x_1})^2 \sum_{s=1}^{S}(x_{2,s}^* - \overline{x_2^*})^2}}, \tag{9}$$

where $x_1$ and $x_2$ are the complex images being analyzed. S is the corresponding pixel number. $\overline{\{\cdot\}}$ represents the mean value operation. $\{\cdot\}^*$ is the complex conjugate operator. The magnitude of cPCC describes the strength of the linear similarity between the two input images, and its phase angle describes the average correlation direction difference of the two images.

**Funding.** Engineering and Physical Sciences Research Council; U.S. Department of Energy (DESC0012704).

**Acknowledge.** The work at Brookhaven National Laboratory (BNL) was supported by the U.S. Department of Energy (DOE), Office of Basic Energy Sciences (BES). The X-ray measurements were performed at the 3-ID Hard X-ray Nanoprobe (HXN) beamline of the National Synchrotron Light Source II (NSLSII), a U.S. DOE BES User Facility operated by Brookhaven National Laboratory under Contract No. DE-SC0012704. The work at UCL was supported by EPSRC.

**Disclosures.** The authors declare no conflicts of interest.

**Data availability.** Data underlying the results presented in this paper are not publicly available at this time but may be obtained from the authors upon reasonable request.

**Supplemental document.** See Supplement 1 for supporting content.